\documentclass[10pt,letterpaper]{article}
\usepackage{opex3}
\usepackage{braket}
\usepackage{amsmath}
\usepackage{gensymb}
\usepackage{soul}

\DeclareMathOperator{\cov}{cov}
\DeclareMathOperator{\corr}{corr}

\begin{document}

\title{Tracking the polarisation state of light via Hong-Ou-Mandel interferometry}

\author{Natapon Harnchaiwat, Feng Zhu, Niclas Westerberg, Erik Gauger, and Jonathan Leach$^*$}

\address{School of Engineering and Physical Sciences, Heriot-Watt University, Edinburgh, EH14 4AS, UK}

\email{$^*$j.leach@hw.ac.uk} 



\begin{abstract}
We provide a statistically robust and accurate framework to measure and track the polarisation state of light employing Hong-Ou-Mandel interference.  This is achieved by combining the concepts of maximum likelihood estimation and Fisher information applied to photon detection events. Such an approach ensures that the Cram\'er-Rao bound is saturated and changes to the polarisation state are established in an optimal manner.  Using this method, we show that changes in the linear polarisation state can be measured with 0.6 arcminute precision (0.01 degrees).
\end{abstract}





\section{Introduction}

Hong-Ou-Mandel (HOM) interference plays a central role in quantum science \cite{walborn2003multimode,lopes2015atomic,kobayashi2016frequency}. This two-photon quantum interference process is the fundamental building block for many quantum computing and quantum information protocols -- it is the key behind quantum networks and quantum repeaters \cite{toyoda2015hong, o2003demonstration, bouwmeester1997experimental, pan1998experimental, pan2003experimental, lo2012measurement, tang2014experimental,chrzanowski2014measurement, lucamarini2018overcoming}. Since its discovery, Hong-Ou-Mandel interference has also been widely used for ultra-sensitive measurements in quantum metrology \cite{ou1988observation,nagata2007beating,xie2015harnessing}. 

In short, Hong-Ou-Mandel interference is the phenomenon where two photons that are incident on two orthogonal input ports of a beam splitter preferentially exit in the same mode \cite{hong1987measurement}. This bunching of photons leads to a reduction of  coincident detection events between the two output ports. For pairs of single photons, how far the rate  of coincident detection events drops depends on the overlap between the two quantum states, and only perfect indistinguishablility in all degrees of freedom of the photons leads to a fall all the way to zero coincidences. Specifically, this includes simultaneous arrival time, as well as perfectly matched spatial, frequency, and polarisation properties. As such, deviation from perfect Hong-Ou-Mandel interference is a reliable and simple method to probe the difference between two pure quantum states.

As quantum interference via Hong-Ou-Mandel interference plays such a crucial role in quantum science, it is vital to understand how changes to the quantum states involved manifest themselves in the observed measurements.  It has recently been shown that attosecond temporal delays are measurable using HOM interference \cite{lyons2018attosecond}. That study provided an accurate statistical framework to measure temporal delays by combining Fisher information and maximum likelihood estimation.  In the work presented in this paper, we apply the same statistical approach to the polarisation degree of freedom. Specifically, we derive the Fisher information for the difference in polarisation between the two photons, and construct a corresponding maximum likelihood estimator. This approach allows us to resolve changes to the linear polarisation state of a photon of order 0.01 degrees. 

\section{Theory}

\subsection{Saturating the Cram\'er-Rao bound}

The goal of this work is to establish and demonstrate an accurate statistical framework for the detection of changes to the polarisation state of light utilising Hong-Ou-Mandel interference. We aim for our approach to saturate the Cram\'{e}r-Rao bound, which is a fundamental lower bound on the variance of the estimated polarisation,
\begin{equation}\label{sup_Cramer}
    {\rm Var}(\widetilde{\Omega}) \geq \frac{1}{NF} ~,
\end{equation}
where $F$ is the Fisher information and $N$ the number of observations $N$. The Fisher information is generally defined as 
\begin{align}
   F_{\Omega} = \sum_i \frac{[\partial_{\Omega} P(i | \Omega)  ]^2}{P( i | \Omega)}.
\end{align}
Here, the conditional probability distribution $P(i | \Omega)$ is the probability that we observe the event $i$ given the parameter $\Omega$. In our case $\Omega$ will be the polarisation difference between the photons. The events that we observe in this experiment are the single and two-photon coincidences (contained within $i$), given the polarisation states of the two photons (contained within $\Omega$). Fisher information has recently been used in quantum science to demonstrate the violation of the shot-noise limit in photonic quantum metrology \cite{Slussarenko}.

In \cite{lyons2018attosecond}, Lyons {\it et al.}~both saturating the inequality \eqref{sup_Cramer} and achieving a small variance for our inferred polarisation can be achieved by combining three independent approaches: (i) using an approach that maximises the Fisher information; (ii) using maximum likelihood estimation to make the best use of information gained in any measurement; (iii) and implementing a measurement protocol that minimises the impact of drift and instability on the experiment data. 

In the following, we will first give a  theoretical description of HOM interference for the polarisation degree of freedom, before  addressing and then combining each of the above-listed approaches.

\subsection{Hong-Ou-Mandel interference for polarisation states }

\begin{figure}[]
\centering
\includegraphics[width=12cm]{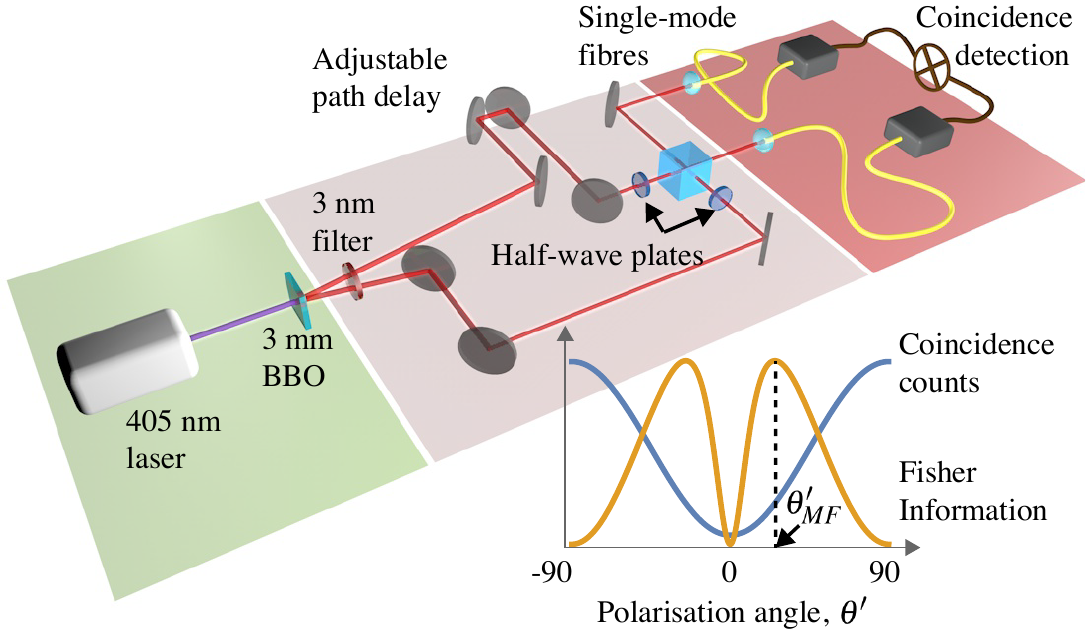}
\caption{Schematic of experiment. A 405 nm laser pumps a 3 mm BBO type I crystal to produce pairs of downconverted photons. We use a 3 nm bandpass filter to spectrally filter the detected modes.  The two photons each pass through a half-wave plate to control their polarisation states.  The photons are incident on a beam splitter, and then single-mode fibres.  Single-photon detectors and coincidence logic records the rate of the single channel and coincidence events.  The inset shows the predicted coincidence counts and associated Fisher information as a function of the linear polarisation angle of one of the photons.}
\label{setup}
\end{figure}

The polarisation state of a pure photon can be represented using the horizontal/vertical basis $\{\ket{H}, \ket{V}\}$. A general polarisation state can then be written using two angles $\theta$ and $\phi$ from the Poincar\'e sphere representation as follows 
\begin{equation}\label{sup_bloch1}
\ket{\psi} = \cos\frac{\theta}{2}\ket{H} + \sin\frac{\theta}{2} e^{i\phi}\ket{V}.
\end{equation}
In a HOM experiment, the probability that two photons in polarisation states $\ket{\psi_a}$ and $\ket{\psi_b}$ incident on the two input modes of a beam splitter will exit in the same output port depends on their indistinguishability. Assuming the photons are identical in all other degrees of freedom, the probability that a coincidence will be observed after the beam splitter is simply given by $P_c = \frac{1}{2}(1-|\braket{\psi_a|\psi_b}|^2)$.  In this work we account for any additional mismatch in other degrees of freedom, i.e.~spatial mode, frequency or time delay, by using $\alpha \in [0,1]$ to represent the indistinguishability of the two photons in all other domains ($\alpha$ = 1 means the two photons are perfectly indistinguishable in the other domains; 0 means the two photons are completely distinguishable in at least one other domain).  The coincidence probability in the context of the polarisation degree of freedom is then
\begin{align}
    P_c = \frac{1}{2}\bigg(1 -\alpha\bigg[&\cos^2\frac{\theta_1}{2}\cos^2\frac{\theta_2}{2} + \sin^2\frac{\theta_1}{2}\sin^2\frac{\theta_2}{2}\nonumber\\ 
    & \quad + \; 2\cos\frac{\theta_1}{2}\cos\frac{\theta_2}{2}\sin\frac{\theta_1}{2}\sin\frac{\theta_2}{2}\cos(\phi_1-\phi_2)\bigg]\bigg),
    \label{sup_bloch3}
\end{align}
where $\theta_1, \phi_1$ and $\theta_2, \phi_2$ are the angles associated with the Poincar\'e spheres for the respective photons.  For the case that the two photons are linearly polarised ($\phi_1 = \phi_2 = 0$), and once we set $\theta_2 = 0$ the coincidence probability simplifies to 
\begin{equation}\label{Pcsimple}
    P_c = \frac{1}{2}\left(1-\alpha\cos^2\frac{ \theta_1}{2}\right).
\end{equation}
The inset of figure \ref{setup} shows how the coincidence probability changes as a function of the linear polarisation state of one of the photons. 

In any realistic experiment, there is always loss, which we capture with a loss rate $\gamma$. Loss can come from, for example, imperfect coupling efficiency or non-unity detector efficiency. For each incoming photon pair, there are three possible outcomes: both detectors click $P_2$, one of detectors clicks $P_1$, and no detectors click $P_0$. These probabilities are related to the coincidence probability  $P_c$ and the probability that two photons bunch together, $P_b = 1- P_c$, as follows \cite{lyons2018attosecond}: 
\begin{align}\label{sub_loss2}\left(
\begin{array}{c}
P_0\\
P_1\\
P_2
\end{array}\right)=\left(\begin{array}{cc}
\gamma^2 & \gamma^2 \\
2\gamma(1-\gamma) & 1-\gamma^2\\
1-2\gamma(1-\gamma)-\gamma^2 & 0
\end{array}\right)\left(\begin{array}{c}
P_c\\
P_b\\
\end{array}\right) = \left(\begin{array}{c}
\gamma^2\\
(1-\gamma^2)-(1-\gamma)^2 P_c\\
(1-\gamma)^2 P_c
\end{array}\right).
\end{align}
%

\subsection{Fisher Information}
As in the discussion above, we consider two photons in polarisation states $\ket{\psi_a}$ and $\ket{\psi_b}$ that are incident on the orthogonal inputs modes of a 50:50 beam splitter, and with two detectors placed after the beam splitter, one in each orthogonal output mode. The events that we can observe are that one or two detectors click; and the probabilities of these events depend on the parameters $\theta_1, \theta_2, \phi_1,$ and $\phi_2$, as well as the visibility $\alpha$ and loss rate $\gamma$.  We therefore look at the Fisher information of $P(1|\theta_1, \theta_2, \phi_1,\phi_2,\alpha,\gamma)$ and $P(2|\theta_1, \theta_2, \phi_1,\phi_2,\alpha,\gamma)$  in order to maximise the information gained from the measurements. 

In our experiment, we choose to measure the difference in linear polarisation state of two photons. As stated above, this allows us to set $\phi_1 = \phi_2 = 0$. Under this condition, it is straightforward to show that the Fisher information depends only on $\theta_1 - \theta_2  = \Delta \theta$. For simplicity, we fix one photon state by setting $\theta_2 = 0$ and change only $\theta_1$. Furthermore, it is from now on convenient to define $\theta^\prime$ as the angle half way between two the polarisation states, i.e. $\theta^\prime = \frac{\theta_1}{2}$. Therefore, we need to consider only the Fisher information with regards to $\theta'$, $\alpha$ and $\gamma$.

Let $\Vec{A}$ be the vector formed by these parameters $[\theta',\alpha,\gamma]$ and $k$ be the number of detectors that click. The elements of the Fisher information matrix are then given by
\begin{equation}\label{sup_Fisher1}
    {\begin{array}{c}
    F_{i,j}(\Vec{A})
    \end{array}} =
    \sum\limits_{k = 0,1,2}P(k|\Vec{A})\left(\frac{\partial}{\partial A_i}\log P(k|\Vec{A}) \right)\left(\frac{\partial}{\partial A_j}\log P(k|\Vec{A})\right),
\end{equation}
and the resulting matrix evaluates to
\begin{equation}\label{sup_Fisher2}
    \begin{split}
        \mathbf{F} &= \kappa\left[\begin{array}{ccc}
          \sin^2\left(2\theta'\right) & -\left(\sin\left(2\theta'\right)\cos^2\theta'\right)/\alpha & -\alpha \chi \sin\left(2\theta'\right) \\
          -\left(\sin\left(2\theta'\right)\cos^2\theta'\right)/\alpha & \left(\cos^4\theta'\right)/\alpha^2 & \chi \cos^2\theta' \\
          -\alpha \chi \sin\left(2\theta'\right)  & \chi \cos^2\theta' & 4 \chi/\left(1-\gamma\right)^2
        \end{array}\right] ,\\
        &\text{ with }\quad \kappa\; = \; \frac{(1-\gamma)^2(1+\gamma)\alpha^2}{(1-\alpha\cos^2\theta')(1+3\gamma+(1-\gamma)\alpha\cos^2\theta')},\\
        &\text{ and }\quad\; \chi\; = \;\left[2 \left(1-\alpha  \cos ^2\theta'\right)\right]/\left[\alpha^2 \left(1-\gamma^2\right)\right].
    \end{split}
\end{equation}
As we are interested in measuring the polarisation angle $\theta'$, we will mostly concern ourselves with the Fisher information related to $\theta'$. This is given by
\begin{equation}
    F_{1,1} = \frac{(1-\gamma)^2(1+\gamma)\alpha^2\sin^2\left(2\theta'\right)}{(1-\alpha\cos^2\theta')(1+3\gamma+(1-\gamma)\alpha\cos^2\theta')}.
\label{eq:fitheta}
\end{equation}

Instead of performing multiparameter estimation, we will here treat $\gamma$ and $\alpha$ as calibration parameters which are to be determined in an experimental pre-stage, similarly to \cite{lyons2018attosecond}. In this calibration process we assign estimated values to $\gamma$ and $\alpha$ according to
\begin{align}
\gamma &= \frac{N_1-N_2}{N_1+3N_2}\bigg|_{\theta' = \pm\pi/2},\nonumber\\
\alpha &= 1-\frac{\min(N_2)}{\max(N_2)},
\end{align}
where $N_1$ and $N_2$ are the observed detector rates that one detector or two detectors click, respectively. The definition of $\alpha$ is straightforward, and $\gamma$ is determined by the ratio between $P_2$ and $P_1$ outside of the dip, i.e.~for a polarisation angle difference $\pi/2$.

We note that imprecision or uncertainty in these calibration parameters imposes a lower bound on the variance of $\theta'$ (which will be necessarily higher than that given by the inverse of Eq.~\eqref{eq:fitheta} times $N$). More specifically, the calibration parameters and $\theta'$ are related  through the covariances
\begin{align}
\cov(A_i,A_j) \geq \frac{(F)^{-1}_{i,j}}{N},
\end{align}
where $A_i$ is the $i^\text{th}$ component of the vector $\vec{A} = \left[\theta',\alpha,\gamma\right]$, and $\mathbf{F}^{-1}$ is the inverse of Eq.~\eqref{sup_Fisher2}. We note that the Fisher information matrix in Eq.~\eqref{sup_Fisher2} is  singular, signifying that there exists a parameter combination which does not alter the coincidence rate to linear order. In particular, 
\begin{align}
\vec{V} = \left[1/(2\alpha),\tan\theta',0\right]
\end{align}
is an eigenvector of $\mathbf{F}$ with zero eigenvalue. Therefore, it is possible to alter the visibility $\alpha$ and the polarisation angle $\theta'$ simultaneously in such a way that the coincidence rate is unaltered, i.e. they exactly counter-balance their influence on $P_2$. Extra care must thus be taken when estimating the visibility $\alpha$, as the uncertainly in $\alpha$ may thus have a large impact on the estimate of $\theta'$. 

With this in mind, let us now form a non-singular Fisher information matrix by restricting $\vec{A}=\left[\theta',\gamma\right]$ in the calculation of $\mathbf{F}$ in Eq.~\eqref{sup_Fisher1}. From this we can see that the uncertainty in $\gamma$ has a more limited impact, and it can be shown that the correlation between those two parameters is $|\corr(\theta',\gamma)|~=~\left|\cov(\theta',\gamma)/\sqrt{\sigma^2_{\theta'}\sigma^2_\gamma}\right|~\leq \left|F^{-1}_{\theta'\gamma}/\sqrt{F^{-1}_{\theta'\theta'}F^{-1}_{\gamma\gamma}}\right| \leq ~0.153 (\alpha = 0.79, \gamma = 0.91$).We find that this quantity is less than 0.153 for any  $\theta$ other than $\pi/2$. Finally, as a part of the estimation procedure, we estimate the number of photons without loss using 
\begin{align}
N = \frac{N_1+N_2}{1-\gamma^2}
\end{align}
when calculating the variance for $\theta'$ according to Eq.~\eqref{sup_Cramer}.

Detector dark counts and after pulsing can result in `false' coincidences: the most likely event is a dark count coalescing with a genuine photon detection in the other detector. This leads to a constant rate of coincidences that is independent of the polarisation, thus reducing the achievable visibility and Fisher information. This is automatically included in our extracted visibility, and we do not account for it separately. In our experiment, significantly fewer than 1~\% of coincidences arise in this way, and this is not a limiting factor to the attainable precision. 

\subsection{Maximum Likelihood Estimation}

In the experiment, we observe detector rates $N_1$ and $N_2$, which are proportional to $P_1$ and $P_2$ respectively.  $N_0$, $N_1$ and $N_2$ are the number of times that no detectors click, one detector clicks and two detectors click, respectively. The probability of observing these outcomes is $P(N_0,N_1,N_2|\theta',\alpha,\gamma)=P_0^{N_0}P_1^{N_1}P_2^{N_2}$, which yields the likelihood function $\mathcal{L}(\theta',\alpha,\gamma|N_0,N_1,N_2)$. 

In order to estimate $\theta'$, we use the maximum of the log-likelihood function. The estimator is found by solving $ \partial_{\theta'}(\log \mathcal{L}) = 0$ for ${\theta'}$, which implies the condition $N_1 P_2 = N_2 P_1$. This in turn yields
\begin{equation}
    \widetilde{\theta'} = \pm\ \arccos\left(\sqrt{\frac{N_1-N_2\left(\frac{1+3\gamma}{1-\gamma}\right)}{\alpha\left(N_1+N_2\right)}}\right).
\end{equation}
In practice, the estimator can yield complex values in a noisy setup. In order to ensure that the estimator yields only real values, we limit the  minimum and maximum polarisation angles to be $-\frac{\pi}{2}$ and $\frac{\pi}{2}$ respectively, and also map complex angles to real angles through two sequential mappings which keep `allowed' values untouched but modify estimates that are out of range:
\begin{equation}
\widetilde{\theta^\prime}_{MAP} = 
\begin{cases}
      \widetilde{\theta^\prime}, & \text{if}\ \ N_1-N_2\left(\frac{1+3\gamma}{1-\gamma}\right) \geq 0 \\
      \pm\frac{\pi}{2}, & \text{if}\ \ N_1-N_2\left(\frac{1+3\gamma}{1-\gamma}\right) < 0 \\
    \end{cases},
\end{equation}
\begin{equation}
\widetilde{\theta^\prime}_{MAP} = 
\begin{cases}
      \widetilde{\theta^\prime}, & \text{if}\ \ \sqrt{\frac{N_1-N_2\left(\frac{1+3\gamma}{1-\gamma}\right)}{\alpha\left(N_1+N_2\right)}} \subset [-1,1] \\
      0, & \text{if}\ \ \sqrt{\frac{N_1-N_2\left(\frac{1+3\gamma}{1-\gamma}\right)}{\alpha\left(N_1+N_2\right)}} \not\subset [-1,1] \\
    \end{cases}.
\end{equation}

\subsection{Allan variance}

\begin{figure}[]
\centering
\includegraphics[]{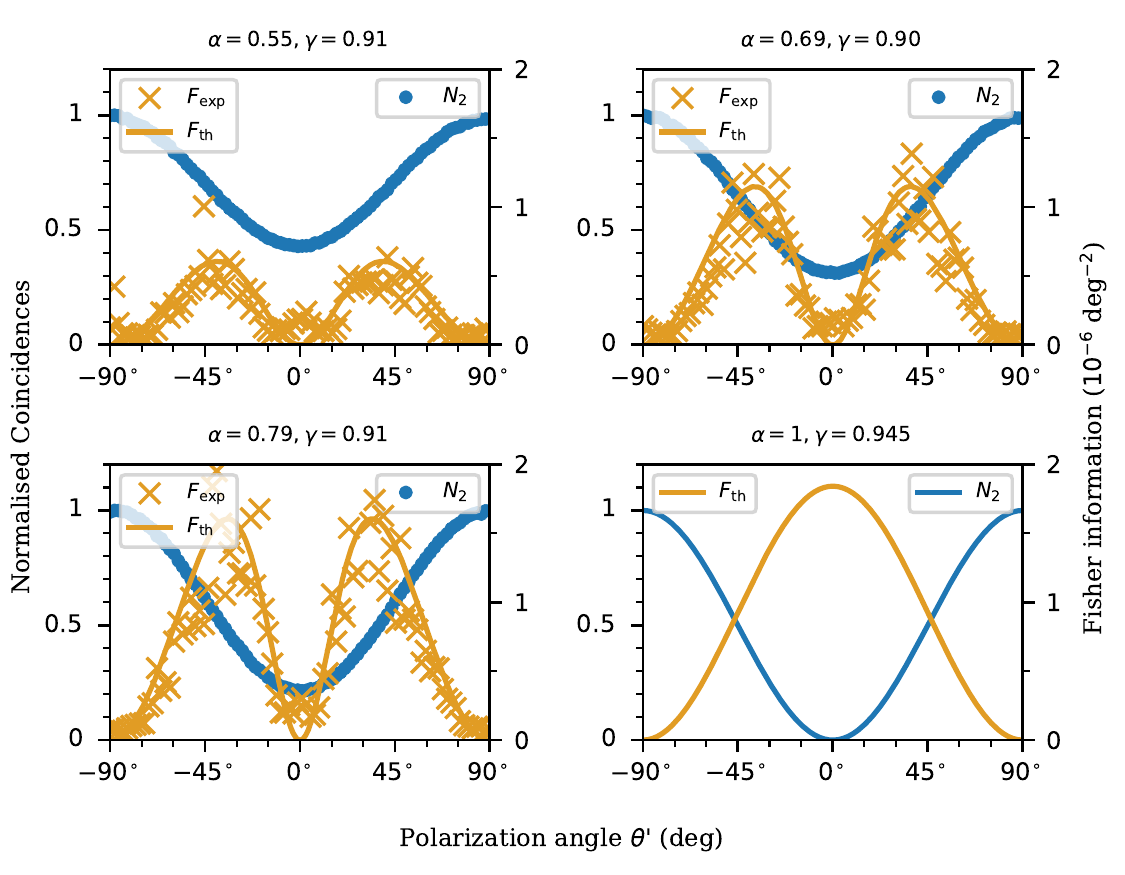}
\caption{Normalised coincidences and inverse variance of the estimator $\widetilde{\theta_1^\prime}$ compared with the predicted Fisher information.  (Top left) $\alpha = 0.55$ and $\gamma = 0.91$; (Top right) $\alpha = 0.69$ and $\gamma = 0.90$; (Bottom left) $\alpha = 0.79$ and $\gamma = 0.91$; (Bottom right) theoretical predictions associated with $\alpha = 1$ and $\gamma = 0.945$.  This data shows no dip at $0^\circ$ as $\alpha = 1$.  The polarisation angle associated with the maximal Fisher information changes according to the parameters $\alpha$ and $\gamma$.  For $\alpha = 0.79$ and $\gamma = 0.91$, $\theta^\prime_{MF} = \pm 34\deg$. The coincidence counts are normalised so that the maximum count is equal to one, and no background subtraction is applied.}
\label{FisherInfo}
\end{figure}
The Allan variance is a mathematical tool to estimate noise in the system in a signal having drift where the standard deviation does not converge \cite{allan1966statistics}. The Allan variance considers the difference in consecutive data points, thereby reducing the effect of the drift on the noise estimate. The Allan variance is here given by
\begin{equation}\label{allan}
    (\sigma_{\theta'}^{Al}(\tau))^2 = \frac{1}{2}\langle(\bar{\theta}'_{n+1}-\bar{\theta}'_{n})^2\rangle ,
\end{equation}
where $\tau$ is the observation period and $\bar{\theta}'_n$ is the average of $\theta'$ in the $n$-th observation period $\tau$. In our experiment, we use the Allan deviation $\sigma^{Al}_{\theta'}(\tau) =  \sqrt{(\sigma^{Al}_{\theta'}(\tau))^2}$ to mitigate against the effect of drift in estimated polarisation angle $\Delta\tilde{\theta}^\prime$. 

In the case that there is only white noise and no long-term drift, the Allan deviation is equal to the standard deviation. To see this, suppose that $\Theta(n)$ is the drift in the system, where the $n$ denotes the $n^\text{th}$ measurement. We can then separate the contributions to the estimate of $\theta'$ into the drift [$\Theta(n)$] and the remainder ($\xi_n$), such that $\theta'_n = \xi_n + \Theta(n)$. In the experiment, the drift $\Theta(n)$ is a monotonically increasing function of $n$, and is approximately linear. Thus we can define $\Theta(n+1)-\Theta(n) = \Delta \Theta$. Now the standard deviation of $\theta'$ is simply square root of the addition of two variances, i.e.
\begin{equation}\label{deviation}
    \sigma_{\theta'} = \sqrt{\sigma_\xi^2 + \sigma_\Theta^2}
\end{equation}
On the other hand, using the Allan deviation from Eq.~\eqref{allan}, we have
\begin{align}\label{allanDeviation}
\sigma^{Al}_{\theta'} &= \sqrt{\frac{1}{2}\langle(\theta'_{n+1} - \theta'_{n})^2\rangle} = \sqrt{\frac{1}{2}\langle(\xi_{n+1} - \xi_{n} + \Delta \Theta)^2\rangle} \nonumber\\ 
&= \sqrt{\frac{1}{2}(2\langle \xi^2\rangle - 2\langle \xi\rangle^2 + \Delta \Theta^2)} = \sqrt{\sigma_\xi^2 + \frac{\Delta \Theta^2}{2}},
\end{align}
where the second line follows from the independence of $\xi_n$ and $\xi_{n+1}$. Now the standard deviation in Eq.~\eqref{deviation} increases with the number of measurements $N$, as the drift is approximately linear and consequently $\sigma_\Theta$ increases. However, since $\Delta \Theta$ is constant in time and is independent of the number of measurements, we find that the Allan deviation in Eq.~\eqref{allanDeviation} does not suffer from this.

\section{Experimental setup and Methods}

\begin{figure}[]
\centering
\includegraphics[]{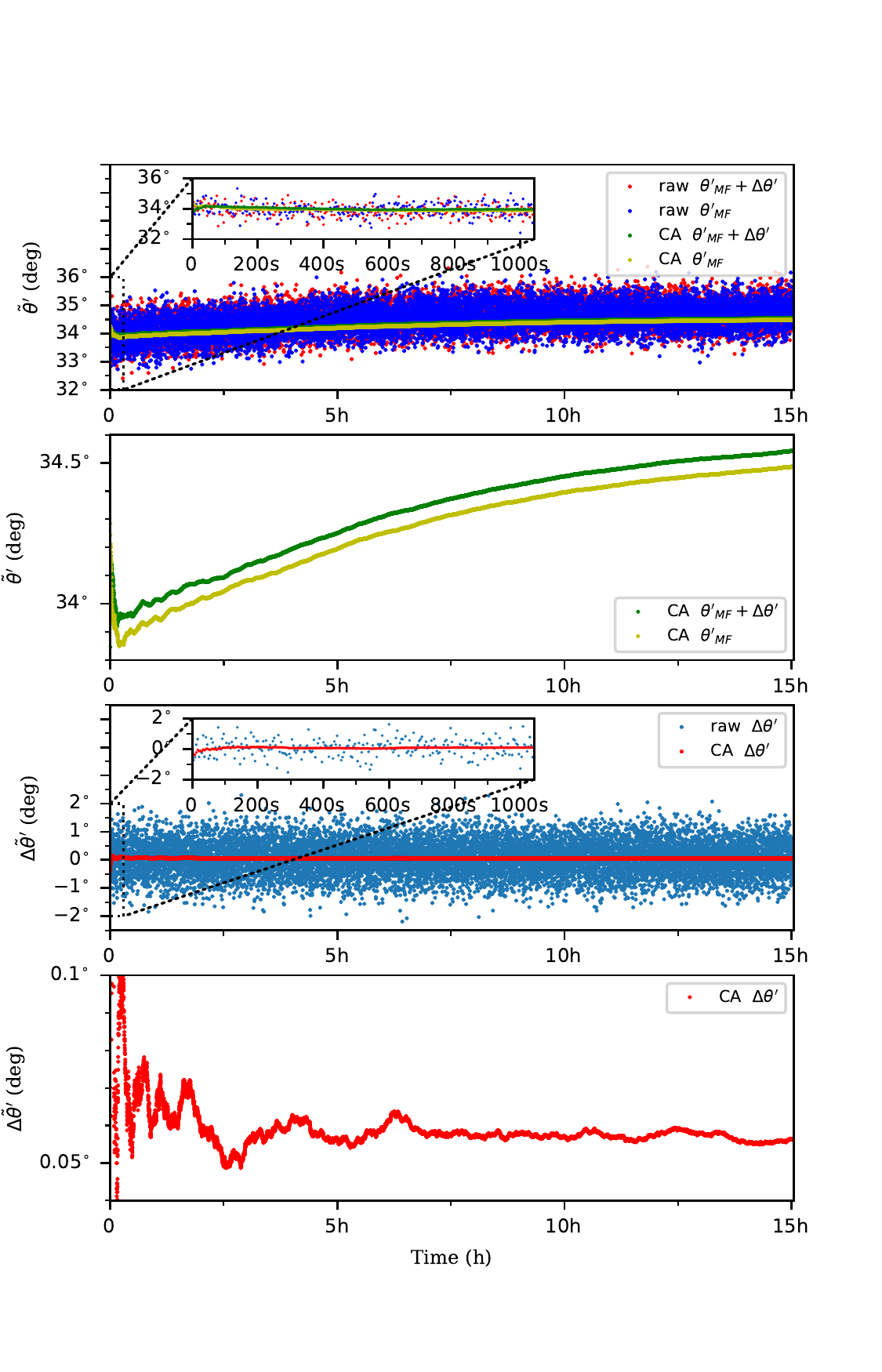}
\caption{(Top) Estimators at the maximum Fisher information angle $\theta^\prime_{MF}$ and at  $\theta^\prime_{MF}+ \Delta\theta^\prime$. (Bottom) The estimated $\Delta \widetilde{\theta^\prime}$, which calculated from the values above. The long-term drift can be mitigated by using the difference approach.}
\label{measurements}
\end{figure}

\begin{figure}[]
\centering
\includegraphics[]{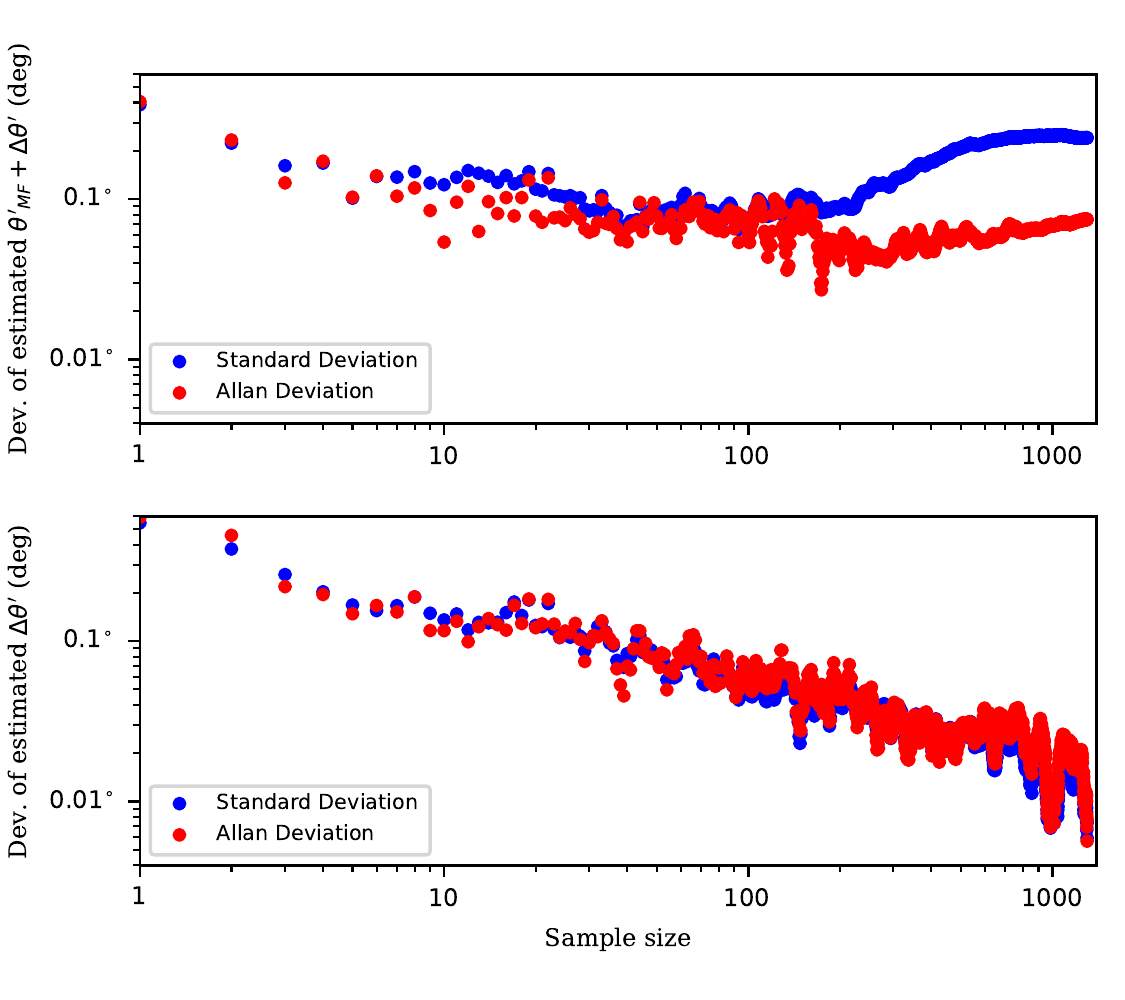}
\caption{Standard and Allan deviations of the estimator for various sample sizes. (Top) The deviation of the estimated $\theta^\prime_{MF} + \Delta \theta^\prime$. (Bottom) The deviation of the estimated $\Delta \theta^\prime$.}
\label{SDAD}
\end{figure}

\begin{figure}[]
\centering
\includegraphics[]{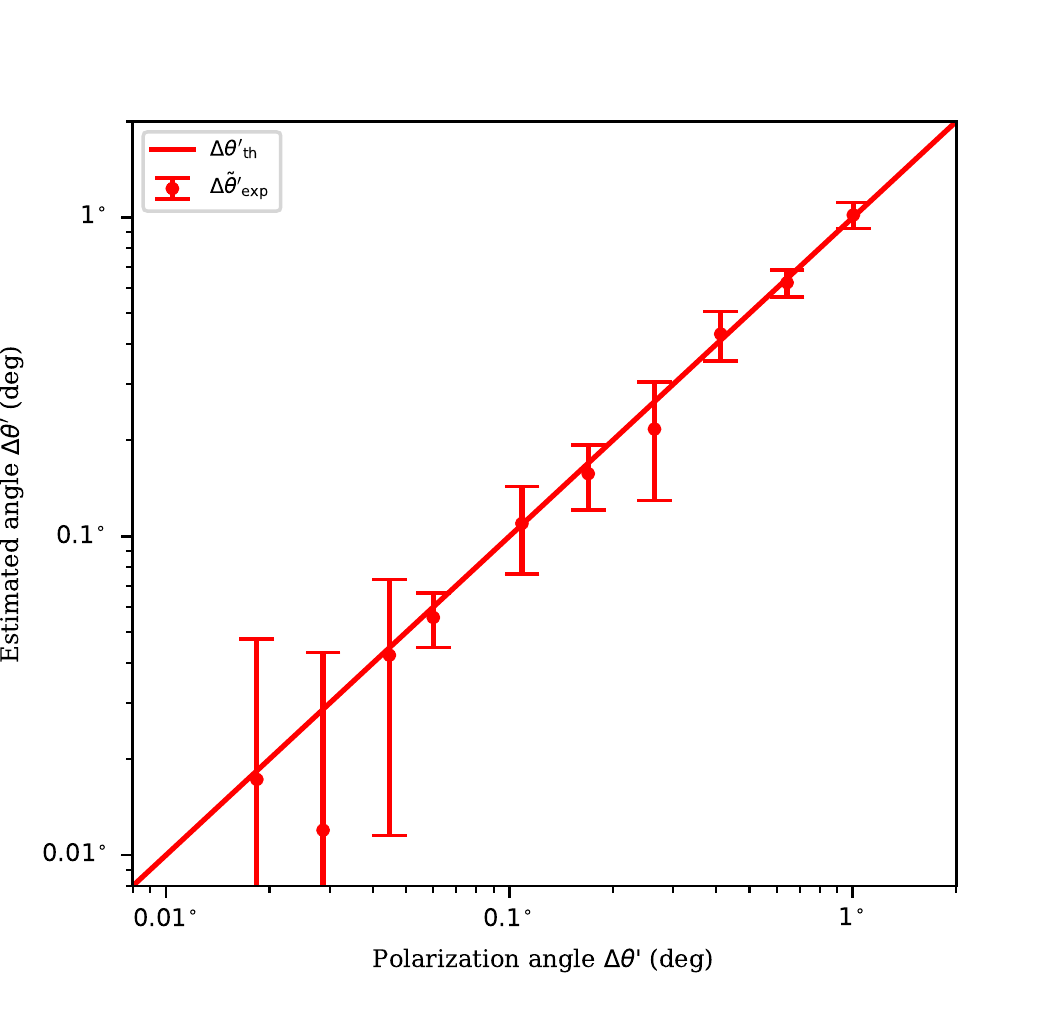}
\caption{The actual change to the polarisation state $\Delta \theta^\prime$ vs.~the estimator $\Delta \widetilde{\theta^\prime}$. These data are plotted on a log-log scale.  We see an accurate measurement of $\Delta \theta^\prime$ over the range 0.01 degree to 1 degree. }
\label{Accuracy}
\end{figure}

A continuous wave 405-nm laser (Coherent OBIS) is used to pump a 3 mm thick BBO type I crystal. Pump photons generate pairs of signal and idler through spontaneous parametric down-conversion (SPDC). In order to achieve high visibility of the HOM interference, we use 3~nm filter to spectrally filter the photons in the  frequency domain. The two photons are spatially separated by a prism placed in the far-field of the crystal.  We use a 150 mm lens placed 150 mm from the crystal, and the prism is placed 150 mm from the lens. The polarisation state of the photons are controlled with half-wave plates. One of the optical paths has a translation stage to match the time domain of two photons and a motorized rotation mount to control the angle of the half-wave plate.  The two paths are then incident on non-polarising 50:50  beam splitter, and the two output modes are coupled into single-mode fibres using 15 mm lenses. The fibres are connected to single-photon avalanche detectors (SPAD), and coincidence detection is performed with a TimeHarp 260 counting board with a 1 ns coincidence window.  See figure \ref{setup} for a schematic of the experimental setup. For our experiment, we find the values of $N, N_1,$ and $N_2$ are of order 120, 20, and 1.3 kcps, respectively.  The dark-count rates of the two detectors were measured to be 1 and 1.5 kcps. Together with our coincidence gate, this corresponds to an accidental coincidence rate of $\approx 1$ cps.

\section{Results}

The first stage of our procedure is to calibrate the system to establish $\alpha$ and $\gamma$. We maximize the Fisher information by maximising the visibility of the HOM dip.  In this experiment, we achieve a maximum visibility of $\alpha = 0.79$ and $\gamma = 0.91$, and this sets the angle of maximal Fisher information to be around $\pm 34$ degrees, see figure \ref{FisherInfo}. The Fisher information also depends on the angle of the two polarisation states, $\theta^\prime$, so we use the angle that provides the highest Fisher information.  Figure \ref{FisherInfo} shows data for the measured coincidence counts and corresponding Fisher information as a function of the polarisation angle $\theta^\prime$. As can be seen from the data, the angle at which we observe the maximum Fisher information depends on the visibility of the coincidence curve and losses.

We attribute the main contribution to the imperfect $\alpha$ to the mode mismatch at the free-space HOM beam-splitter.  Higher visibilities could be easily achieved with fibre beam-splitters as e.g.~in ref \cite{lyons2018attosecond}, but this would introduce an unknown polarisation state at the point of the HOM interference.  The reduction in $\gamma$ from unity is due to a combination of the detector efficiencies ($\approx 65\%$ at 810 nm) and coupling efficiencies.

The data in figure \ref{FisherInfo} is the inverse of the variance of the estimator multiplied with expected number of photon pairs $N$, and this is compared with the predicted Fisher information. As the inverse of the variance agrees well with the predicted Fisher information, this means that we saturate the Cram\'er-Rao bound. Overall, the inverse variance shows a good fit with theory, however, angles near zero and $\pm 90$ degrees show slightly lower variance than expected. This is because of the symmetry of the HOM dip. The estimator is only able to give the magnitude of polarisation angles but not their signs and the polarisation angle is limited from 0 to 90 degrees. As a result, angles near zero and 90 degrees have a lower inverse variance than the predicted. Nevertheless, this does not effect the high resolution measurement as that is done at the maximum Fisher information angle where the inverse variance agrees with the predicted Fisher information.

In the reality, this experiment is subject to drift. Any drift effects both $\alpha$ and $\gamma$, and the source of this comes mainly from the temperature fluctuations in the lab and long-term stability of the equipment.  These are of order $\pm 0.1 ~\degree$C over a 20 minute period but significant enough to influence the measurements. Even though $\alpha$ and $\gamma$ are subject to small changes over the course of the experiment, our measurement protocol mitigates the effect of this, as we explain in the following.

Our procedure is to first rotate the half-wave plate to the position with the highest Fisher information and acquire data for one second. This provides an estimate of $\theta_{MF}^\prime$. We then rotate the half-wave plate by a small angle $\Delta\theta^\prime$ so that the angle between to polarisation states is $\theta_{MF}^\prime + \Delta\theta^\prime$.  The new angle is estimated by acquiring data for one second, and the change in angle $\Delta \widetilde{\theta^\prime}$ is calculated as the difference between the two measurements. Including the time for the motorized rotation stage to change, it takes about four to five seconds to complete one round of measurement. The frequency of switching between two angles is much faster than the frequency of the drift, and this protocol is repeated many times to achieve the highest precision. 

Figure \ref{measurements} shows a typical data set of the measurements over a 15 hour period. As can be seen, the effects of drift are significantly reduced in the estimated value of $\Delta \widetilde{\theta^\prime}$.  Figure \ref{SDAD} shows the Allan deviation and the standard deviation.  The close agreement between the two shows that the effects of drift in the system have been mitigated by the measurement protocol.   Both standard and Allan deviations decrease to about 0.01 degrees when sample size is about $10^3$. Figure \ref{Accuracy} shows that the standard deviation of the most precise measurement is 0.01 degrees, which corresponds to 0.6 arcminutes. The fact that the difference between the two measured angles remains constant throughout the experiment is evidence that our procedure is not sensitive to any changes in $\alpha$ or $\gamma$, see the second panel of figure \ref{measurements} and the bottom panel in figure \ref{SDAD}.

\section{Discussion and conclusion}

By combining the concepts of Fisher information and maximum-likelihood estimation, we provide a statically accurate and robust framework in which to measure the polarisation state of light in the context of Hong-Ou-Mandel interference.   We provide the full Fisher information matrix for the polarisation state of the two input photons, and we show that using this method, changes to the linear polarisation state of one of the photons can be established to approximately $\pm 0.01$ degrees.  By using maximum likelihood estimation, we have an estimator that saturates the Cram\'er-Rao bound. This is confirmed as the inverse variance of the estimator agrees well with the Fisher information predicted by our theory.

The precision that we achieve is of order $10^5$ times lower than the FWHM of the width of the HOM dip, which is equal to 90 degrees.  This is the same precision relative to the dip width that was achieved in ref \cite{lyons2018attosecond}.  In that work, a 10 nm filter was used, providing a dip width in the spatial degree of freedom of $\approx 64 \mu m$.  The precision on the displacements was measured to be of order a few nanometers, giving around a $10^5$ times improvement.  The consistency between these two results indicates at the need for HOM dips with very narrow widths if more precise measurements are to be achieved. 

 \section*{Acknowledgement}

We acknowledge helpful discussions with Ashley Lyons and Daniele Faccio. This work was supported by EPSRC grants EP/R024170/1 and EP/M01326X/1. EMG acknowledges financial support from the Royal Society of Edinburgh and the Scottish Government.
 
\section{Disclosures}

The authors declare no conflicts of interest.


\begin{thebibliography}{10}
\newcommand{\enquote}[1]{``#1''}



\bibitem{walborn2003multimode}
S.~Walborn, A.~De~Oliveira, S.~P{\'a}dua, and C.~Monken, \enquote{Multimode
  hong-ou-mandel interference,} Physical review letters \textbf{90}, 143601
  (2003).

\bibitem{lopes2015atomic}
R.~Lopes, A.~Imanaliev, A.~Aspect, M.~Cheneau, D.~Boiron, and C.~I. Westbrook,
  \enquote{Atomic hong--ou--mandel experiment,} Nature \textbf{520}, 66 (2015).

\bibitem{kobayashi2016frequency}
T.~Kobayashi, R.~Ikuta, S.~Yasui, S.~Miki, T.~Yamashita, H.~Terai, T.~Yamamoto,
  M.~Koashi, and N.~Imoto, \enquote{Frequency-domain hong--ou--mandel
  interference,} Nature Photonics \textbf{10}, 441 (2016).

\bibitem{toyoda2015hong}
K.~Toyoda, R.~Hiji, A.~Noguchi, and S.~Urabe, \enquote{Hong--ou--mandel
  interference of two phonons in trapped ions,} Nature \textbf{527}, 74 (2015).

\bibitem{o2003demonstration}
J.~L. O'Brien, G.~J. Pryde, A.~G. White, T.~C. Ralph, and D.~Branning,
  \enquote{Demonstration of an all-optical quantum controlled-not gate,} Nature
  \textbf{426}, 264 (2003).

\bibitem{bouwmeester1997experimental}
D.~Bouwmeester, J.-W. Pan, K.~Mattle, M.~Eibl, H.~Weinfurter, and A.~Zeilinger,
  \enquote{Experimental quantum teleportation,} Nature \textbf{390}, 575
  (1997).

\bibitem{pan1998experimental}
J.-W. Pan, D.~Bouwmeester, H.~Weinfurter, and A.~Zeilinger,
  \enquote{Experimental entanglement swapping: entangling photons that never
  interacted,} Physical Review Letters \textbf{80}, 3891 (1998).

\bibitem{pan2003experimental}
J.-W. Pan, S.~Gasparoni, R.~Ursin, G.~Weihs, and A.~Zeilinger,
  \enquote{Experimental entanglement purification of arbitrary unknown states,}
  Nature \textbf{423}, 417 (2003).

\bibitem{lo2012measurement}
H.-K. Lo, M.~Curty, and B.~Qi, \enquote{Measurement-device-independent quantum
  key distribution,} Physical review letters \textbf{108}, 130503 (2012).

\bibitem{tang2014experimental}
Z.~Tang, Z.~Liao, F.~Xu, B.~Qi, L.~Qian, and H.-K. Lo, \enquote{Experimental
  demonstration of polarization encoding measurement-device-independent quantum
  key distribution,} Physical review letters \textbf{112}, 190503 (2014).

\bibitem{chrzanowski2014measurement}
H.~M. Chrzanowski, N.~Walk, S.~M. Assad, J.~Janousek, S.~Hosseini, T.~C. Ralph,
  T.~Symul, and P.~K. Lam, \enquote{Measurement-based noiseless linear
  amplification for quantum communication,} Nature Photonics \textbf{8}, 333
  (2014).

\bibitem{lucamarini2018overcoming}
M.~Lucamarini, Z.~L. Yuan, J.~F. Dynes, and A.~J. Shields, \enquote{Overcoming
  the rate--distance limit of quantum key distribution without quantum
  repeaters,} Nature \textbf{557}, 400 (2018).

\bibitem{ou1988observation}
Z.~Ou and L.~Mandel, \enquote{Observation of spatial quantum beating with
  separated photodetectors,} Physical review letters \textbf{61}, 54 (1988).

\bibitem{nagata2007beating}
T.~Nagata, R.~Okamoto, J.~L. O'Brien, K.~Sasaki, and S.~Takeuchi,
  \enquote{Beating the standard quantum limit with four-entangled photons,}
  Science \textbf{316}, 726--729 (2007).

\bibitem{xie2015harnessing}
Z.~Xie, T.~Zhong, S.~Shrestha, X.~Xu, J.~Liang, Y.-X. Gong, J.~C. Bienfang,
  A.~Restelli, J.~H. Shapiro, F.~N. Wong \emph{et~al.}, \enquote{Harnessing
  high-dimensional hyperentanglement through a biphoton frequency comb,} Nature
  Photonics \textbf{9}, 536 (2015).

\bibitem{hong1987measurement}
C.-K. Hong, Z.-Y. Ou, and L.~Mandel, \enquote{Measurement of subpicosecond time
  intervals between two photons by interference,} Physical review letters
  \textbf{59}, 2044 (1987).

\bibitem{lyons2018attosecond}
A.~Lyons, G.~C. Knee, E.~Bolduc, T.~Roger, J.~Leach, E.~M. Gauger, and
  D.~Faccio, \enquote{Attosecond-resolution hong-ou-mandel interferometry,}
  Science advances \textbf{4}, eaap9416 (2018).

  \bibitem{Slussarenko}
S.~Slussarenko, M.~M.~Weston, H.~M.~Chrzanowski, L.~K.Shalm, V.~B.~Verma, S.~W.~Nam, and G. J. Pryde, \enquote{Unconditional violation of the shot-noise limit in photonic quantum metrology,} Nature Photonics  \textbf{11}, 700–703 (2017).


\bibitem{allan1966statistics}
D.~W. Allan, \enquote{Statistics of atomic frequency standards,} Proceedings of
  the IEEE \textbf{54}, 221--230 (1966).
  

\end{thebibliography}
\end{document}